# FPGA Implementation of High Speed Baugh-Wooley Multiplier using Decomposition Logic

Ananda Kiran[1] and Navdeep Prashar[2]

[1]Department of Electronics and Communication Engineering, Bahra University, Shimla-Hills, H.P, India

*kiranananda7@gmail.com*

[2]Assistant Professor in the Department of Electronics and Communication Engineering, Bahra University, Shimla-Hills, H.P., India

*nav.prashar@gmail.com*

## *Abstract*

*The Baugh-Wooley algorithm is a well-known iterative algorithm for performing multiplication in digital signal processing applications. Decomposition logic is used with Baugh-Wooley algorithm to enhance the speed and to reduce the critical path delay. In this paper a high speed multiplier is designed and implemented using decomposition logic and Baugh-Wooley algorithm. The result is compared with booth multiplier. FPGA based architecture is presented and design has been implemented using Xilinx 12.3 device.*

## 1. Introduction

Multipliers play a pivotal role in many high performance systems such as Microprocessor, FIR filters, Digital Processors, etc. In its early stage, multiplication algorithms were proposed by Burton and Noaks in the year 1968, by Hoffman in the year 1986 and by Guilt and De Mori in the year 1969 for positive numbers. In the year of 1973 and 1979, Baugh-Wooley and Hwang proposed multiplication algorithm for numbers in two's complement form. Multiplication is hardware intensive and the main criteria of interest are higher speed, lower cost and lower power [1]. With development in technology, several researchers have tried multipliers which provide design targets such as low power consumption, increased speed, and regularity of layout or combination of them in one multiplier. This helps making them suitable for achieving compact high speed and low power implementation.

The performance of a system is generally controlled by the performance of the multiplier as the multiplier is usually the slowest element in the system. Furthermore, multiplier is normally the most area consuming element in the system. Therefore, optimizing its speed and area are vital design factors. However, area and speed are generally the conflicting constraints improving speed which results mostly in large area.

With ever increasing applications in portable equipment and mobile communications, the demand for high performance, low-power VLSI systems is gradually increasing. Digital signal processors and application specific integrated circuits depend on the efficient implementation of arithmetic circuits (adder and multiplier) to execute dedicated algorithm such as convolution, correlation and filtering [2]. A Baugh-Wooley multiplier using decomposition logic is presented here which increases speed when compared to the booth multiplier.

## 2. Baugh-Wooley multiplier

In signed multiplication the length of the partial products and the number of partial products will be very high. So an algorithm was introduced for signed multiplication called as Baugh-Wooley algorithm. The Baugh-Wooley multiplication is one amongst the cost-effective ways to handle the sign bits. This method has been developed so as to style regular multipliers, suited to 2's compliment numbers.

### 2.1 Baugh-Wooley Architecture

Baugh-Wooley multiplier hardware architecture is shown in figure 1. It follows left shift algorithm. Mux can select which bit will multiply. Suppose we multiply +4 and -4 in decimal we get '0'. Now, after representing these numbers in two's compliment form we get +4 as 0100 and -4 as 1100. On adding these two binary numbers we get 10000. Discard carry, then number is represented as '0'.

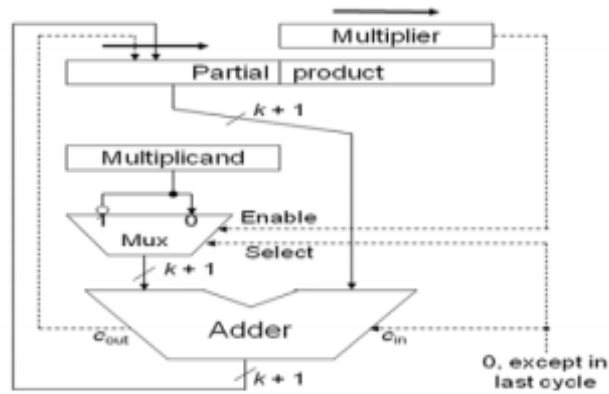

Figure.1 Hardware implementation of Baugh-Wooley Multiplier [3]

Let two n-bit numbers, number (A) and number (B), A and B are often pictured as

$$A = -a_{n-1}2^{n-1} + \sum_{i=0}^{n-2} a_i 2^i \tag{1}$$

$$B = -b_{n-1}2^{n-1} + \sum_{i=0}^{n-2} b_i 2^i \tag{2}$$

Where $a_i$ and $b_i$ area unit the bits during A and B, severally and $a_{n-1}$ and $b_{n-1}$ area unit the sign bits. The full precision product, $P = A \times B$, is provided by the equation:

$$P = A \times B = \left[\left(-a_{n-1}2^{n-1} + \sum_{i=0}^{n-2} a_i 2^i\right) \times \left(-b_{n-1}2^{n-1} + \sum_{i=0}^{n-2} b_i 2^i\right)\right]$$

$$= a_{n-1}b_{n-1}2^{2n-2} + \sum_{i=0}^{n-1} a_i 2^i \sum_{j=0}^{n-2} b_j 2^j - 2^{n-1}\sum_{i=0}^{n-2} a_i b_{n-1} 2^i - 2^{n-1}\sum_{j=0}^{n-2} a_{n-1} b_j 2^j \quad (3)$$

The first two terms of above equation are positive and last two terms are negative [4]. In order to calculate the product, instead of subtracting the last two terms, it is possible to add the opposite values [5]. The above equation signifies the Baugh-Wooley algorithm for multiplication process in two's compliment form.

Baugh-Wooley Multiplier provides a high speed, signed multiplication algorithm [5]. It uses parallel products to complement multiplication and adjusts the partial products to maximize the regularity of multiplication array [6]. When number is represented in two's complement form, sign of the number is embedded in Baugh-Wooley multiplier. This algorithm has the advantage that the sign of the partial product bits are always kept positive so that array addition techniques can be directly employed [6]. In the two's complement multiplication, each partial product bit is the AND of a multiplier bit and a multiplicand bit, and the sign of the partial product bits are positive [6].

## 3. Decomposition Logic

The implementation of digital multiplier with decomposition logic is presented here. In this technique the multiplication process is split into smaller sub-units (smaller multipliers) and their outputs are combined to get the final result, the decomposition logic requires extra circuitry to perform the final addition of outputs attained from the smaller multiplier [7]. However, due to parallel processing, noticeable improvement in speed is achieved.

To check the performance of the multiplier structure, 8×8 multiplier structure is designed using Baugh-Wooley algorithm and the decomposition logic. Fig. 2 [7] shows an 8×8 multiplier implemented using the decomposition logic. In the first stage, four 4×4 multipliers are used to combine all the partial products, the outputs from these 4×4 multipliers are then combined in a treelike fashion to get the final results [7]. The 4×4 multiplier was implemented using Baugh-Wooley method. For 16×16 multiplication, three decomposition structures can be implemented. The first using 4×4 Baugh-Wooley multipliers, the second using 8×8 Baugh-Wooley multipliers and the third using 8×8 decomposition structure [7].

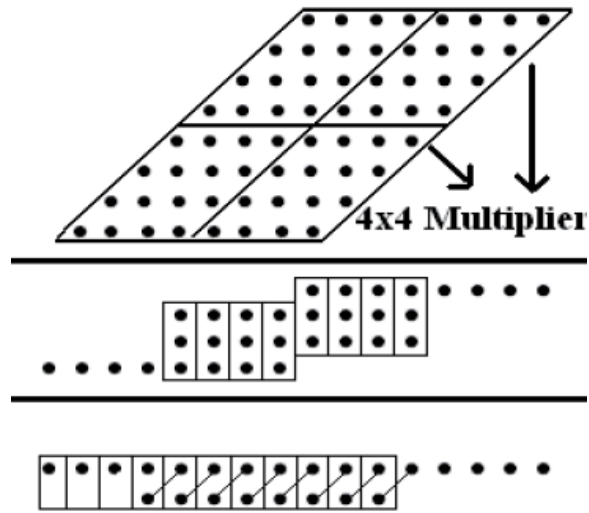

Figure 2. Decomposition structure for 8×8 multiplication

Figure 3 shows the RTL view of Baugh-Wooley Multiplier with decomposition logic.

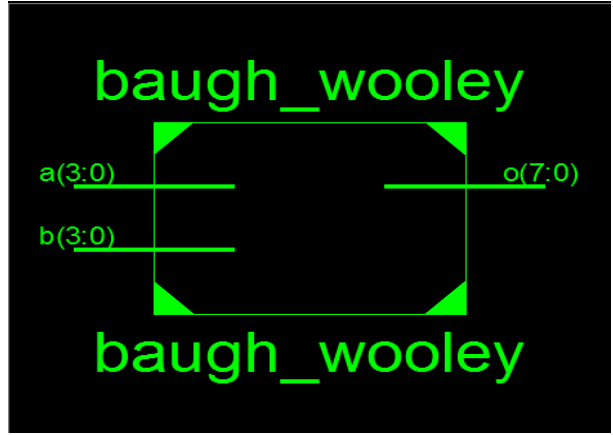

Figure 3. RTL view of Baugh-Wooley with decomposition logic

For any number of inputs one output is generated. Number of bits of the output depends on the number of bits of the inputs; e.g. in figure 3 there are two inputs of 4 bits the output will be of 8 bits.

## 4. Simulation and Result

The code of Baugh-Wooley multiplier and decomposition logic is written in VHDL and simulated using ISim (VHDL/Verilog). The proposed architecture is implemented on Virtex6 xc6vlx75t-3-ff484 device using XILINX 12.3. Table 1 shows the device utilization summary of Baugh-Wooley multiplier using decomposition logic.

Table 1. Hardware Device Utilization Summary

| S. NO. | Logic Utilization | Used | Available |
|---|---|---|---|
| 1 | No. of bonded IO | 16 | 240 |
| 2 | No. of slice LUTs | 13 | 46560 |
| 3 | No. of occupied slices | 6 | 11640 |

### 4.1 Comparison of present work, Baugh-Wooley multiplier and decomposition logic with the previous work

Present work is implemented on Virtex6 xc6vlx75t-3-ff484 device. Thus, finally comparison of present work with previous work is done as shown in Table 2. The simulation results for 8×8 multipliers are summarized in Table 2. For the 8×8 multiplier structure, the Baugh-Wooley method and decomposition logic show an improvement in delay compared to Booth multiplication method due to parallel processing of data.

Table 2. Comparison of present work with the previous reference paper

| 8×8 Multiplier | Booth Multiplier [13] | Baugh-Wooley |
|---|---|---|
| Path Delay | 15.345ns | 10.516ns |
| No. of boned IO | 36 | 32 |
| Total real time to Xst completion | 12.00secs | 6.00secs |
| No. of slice LUTs | 190 | 13 |
| Average fanout | 4.00 | 3.43 |
| Maximum Frequency MHz | 65.16 | 95.09 |

Frm Table 2, it is clear that present design shows an improvement in speed with reduction in used resources on target device.

## 5. Conclusion

In this paper, decomposition logic is implemented with Baugh-Wooley multiplier which shows the better results in terms of path delay and speed. The design operates on maximum frequency of 95.9MHz. The considerable increase in speed make the design suitable for many high performance system such as Digital Signal Processors, FIR filters, Microprocessors etc. The above results show that the Baugh-Wooley Multiplier with decomposition logic is higher in speed as compared to the Booth Multiplier and process the inputs fast to produce the result.

**REFERENCES**

[1] Indrayani Patle, Akansha Bhargav, Prashant Wanjari, "Implementation of Baugh-Wooley Multiplier Based on Soft-Core Processor", IOSR Journal of Engineering (IOSRJEN), Vol. 3, Issue 10, 2013.
[2] Hsin-Lei Lin, Design of a Novel Radix – 4 Booth Multiplier, the 2004 IEEE Asia – Pacific Conference on Circuit and Systems, December 2005
[3] Indrayani Patle, Akanksha Bhargav, Prashant Wanjari, "Implementation of Baugh-Wooley Multiplier Based on Soft-Core Processor", IOSR Journal of Engineering, Vol. 3, Issue 10, Oct. (2013).Based on Soft-Core Processor", IOSR Journal of Engineering, Vol. 3, Issue 10, Oct. (2013).
[4] Jashin Mathews Joseph and V. Sarada, "Reconfigurable High Performance Baugh-Wooley Multiplier for DSP Application", ITSI Transaction on Electrical and Electronics Engineering, Vol. 1, Issue 4 (2013).
[5] C. R. Baugh and B. A. Wooley, "A two's complement parallel array multiplication algorithm," IEEE Trans Comp., vol. C-22, no. 12, pp. 1045-1047, Dec. 1973.
[6] Sudhakar Aswathy and D. Gokila,"High-Speed Power-Efficient Modified Baugh-Wooley Multipliers".
[7] Sundeepkumar agarwal, Palaniappan Ramanathan, Ponnisamy Thangapandian Vanathi" High Speed Multiplier Design Using Decomposition Logic" Serbian Journal of Electrical Engineering Vol. 6, No. 1, 2009 33-42.
[8] MicroBlaze Processor Reference Guide Embedded Development Kit EDK 13.1, www.xilinx.com
[9] C. H. Chang, J. Gu, M. Zhang: A Review of 0.18μm Full Adder Performances for Tree Structured Arithmetic Circuit, *IEEE Transaction on Very Large Integration Systems*, Vol. 13, No. 6, 2005, pp. 686-695.
[10] Y.N. Ching, "Low-Power high-speed multipliers", *IEEE Transactions on Computers*. pp, 1196-1200, Nov. (2005).
[11] Sandeep Shrivastava, Jaikaran Singh and Mukesh Tiwari, "Implementation of Radix-2 Booth Multiplier and Comparision with Radix-4 Encoder Booth Multiplier", SSSITS Sehore, (M.P.), Feb. (2011).
[12] K. Hwang, Computer Arithmetic: Principles, Architecture, and Design. John-Wiley, 1979.
[13] Laxman S. Darshan Prabhu R, Mahesh S Shetty, Mrs Manjula BM, Dr. Chirag Sharma, "FPGA Implementation of Different Multiplier Architectures", International Journal of Different Multiplier Architectures, Vol. 2, Issue 6, June 2012.

Navdeep Prashar and Balwinder Singh, "FPGA Implementation of Pipelined CORDIC Sine Cosine Digital Wave Generator", CS & IT 05, pp. 435–440, © CS & IT-CSCP 2012.

## Authors Biography

**Navdeep Prashar** has obtained his Bachelor of Technology degree from the CT Institute of Engineering, Management and Technology, Jalandhar affiliated to Punjab Technical University, Jalandhar in 2010 and Master of Technology degree from Centre for Development of Advanced Computing (CDAC), Mohali. He is currently serving as Assistant Professor in Bahra University, Waknaghat, Solan, H.P., India. He has 3+ years of teaching experience to both undergraduate and postgraduate students. Mr. Prashar has published one book and presented many papers in the International & National Journals and Conferences. His current interest includes Embedded Systems, VLSI Design & Testing, Low Power techniques, and System on Chip.

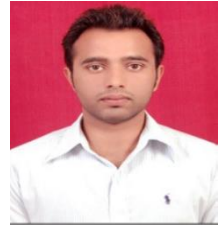

**Ananda Kiran** has obtained her B.Tech. (Electronics and Communication Engineering) degree from the LR Institute of Engineering and Technology, Solan affiliated to Himachal Pradesh University, in 2013, and presently she is doing M.Tech. (Electronics and Communication Engineering) degree from Bahra University, Waknaghat. At present, she is working on her M. Tech. (Electronics and Communication Engineering) thesis. Her area of interest is Embedded Systems, VLSI Design and Digital Signal Processing.

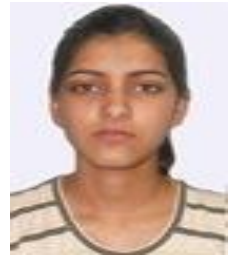